\documentclass[conference]{IEEEtran}
\IEEEoverridecommandlockouts
\usepackage[pdftex]{graphicx}

\usepackage{subfig}

\usepackage{tabularx}
\usepackage{multirow}

\usepackage{gensymb}

\usepackage{stfloats}

\begin{document}
%
\title{A Wised Routing Protocols for Leo Satellite Networks}

\author{\IEEEauthorblockN{Saeid Aghaei Nezhad Firouzja}
\IEEEauthorblockA{College of Electronic Engineering\\ Shanghai Jiang Tong University\\ China\\
Email:aghaeesaeed@ymail.com}
\and
\IEEEauthorblockN{Muhammad Yousefnezhad}
\IEEEauthorblockA{College of Computer Science \\ \& Technology Nanjing University \\ of Aeronautics and Astronautics, China\\
Email:myousefnezhad@nuaa.edu.cn}
\and
\IEEEauthorblockN{Masoud Samadi, Mohd Fauzi Othman$^{\star}$
\thanks{$\star$ Corresponding Author}}
\IEEEauthorblockA{Center for Artificial Intelligence and Robotics\\
Universiti Teknologi Malaysia\\
Kuala Lumpur , Malaysia\\
Email:solariseir@IEEE.org,fauzi@fke.utm.my}}


\maketitle

\begin{abstract}
This Study proposes a routing strategy of combining a packet scheduling with congestion control policy that applied for LEO satellite network with high speed and multiple traffic. It not only ensures the QoS of different traffic, but also can avoid low priority traffic to be "starve" due to their weak resource competitiveness, thus it guarantees the throughput and performance of the network. In the end, we set up a LEO satellite network simulation platform in OPNET to verify the effectiveness of the proposed algorithm.
\end{abstract}

\normalfont\fontsize{9pt}{1.2}{{\textbf{\textit{Keywords-\lowercase{{Satellite Networks; Multi-Service Quality of Service (QoS); Packet Scheduling; Congestion Control; OPNET;}}}}}

%

\IEEEpeerreviewmaketitle

\section{Introduction}
With wide coverage, long transmission distance, independence from geographical location, high spectrum efficiency, large capacity of communication, high communication quality, flexible and quick networking technology, strong ability to resist natural disasters, etc., satellite network enables the communication between any point on earth at any time anywhere, forms beneficial supplement to the ground communication network, realizes global network interconnection and becomes an ideal part of the future Internet [1]. The future Internet will carry a vast scale of traffic with multiple types, thus a QoS routing ensure the quality of service of different traffic is a key technology of future Internet [2]. Additionally, the resource on-board is constrained, load distribution on satellite is unbalanced in terms of time and space, traffic on board is constantly changing with the moving of sub-satellite point, these lead to congestion of some satellite node in the network, due to which network throughput drops [3]. Contra pose different QoS requirements for multi-service and traffic distribution imbalance in network, this paper puts forward a novel compound congestion control routing algorithm to guarantee the QoS requirements of traffic. Traffic is classified into two broad classes—real-time (class A) and non-real-time (class B), bandwidth is allocated reasonably to each type of service according to their QoS requirements and service levels, in the view of differentiated service, when node congestion is detected, non-real-time business with low priority is bypassed to the current relatively idle satellite nodes in order to save more bandwidth for real-time traffic and ensure QoS requirements for each type of traffic.

\subsection{The algorithm description}

\subsubsection{The on-board resource allocation policy}
Traffic supported by the satellite network is developing in the direction of diversification constantly, can be classified into real-time and non-real-time traffic basically. Real-time traffic with the highest service level requires to ensure "three low one guarantee" (low delay, low jitter, low packet loss rate, bandwidth guarantee) QoS [4]; time delay and delay jitter requirement for non-real-time traffic is relatively low, usually use traffic arrival rate as its QoS measure. Satellite, as a relay node of the whole communication network, has limited resources, finding a method which can manage and allocate the limited resource effectively is necessary to assure the QoS of all traffic. 

Packet scheduling is an effective technology to solve resources competition, it decides which packet to send from the waiting queues according to certain rules, which enables input traffic share the output bandwidth in accordance with the predetermined way, is one of core technologies to realize network QoS control. Frequently-used packet scheduling algorithms according to service rules can be divided into simple queue scheduling algorithm(FIFO, PQ, QLT, etc.), scheduling algorithm based on round-robin (RR, WRR and DRR and URR, etc.), scheduling algorithm based on the GPS model (WFQ ,WF2Q ,SPFQ, PFQ) [5].This paper propose a hybrid scheduling algorithm combining PQ and WRR scheduling algorithm-PQWRR algorithm, which set priority queue for traffic class A to ensure its priority; traffic class B is divided into different service level, each service level share the remainder bandwidth using the WRR scheduling policy, ensure the relative fairness of bandwidth allocation. The on-board resource scheduling strategy is shown in Figure \ref{fig:1}:

\begin{figure}[b]
\centering
\includegraphics[width=.7\linewidth]{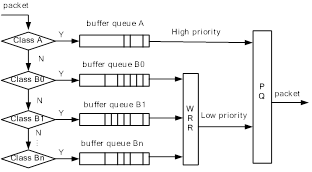}
\caption{On-board resource scheduling strategy (PQWRR)}
\label{fig:1}
\end{figure}

Traffic will be put into different buffer queues depending on its type once it arrives, traffic class A will be transferred to high priority queue and wait to be sent forward, traffic class B will be transferred to corresponding low priority buffer queues, when buffer queue A is empty, traffic class B will be sent forward according to WRR scheduling policies. This strategy combines the advantages of algorithms WRR and PQ, overcomes their both shortcomings and support multi-service traffic better.

\subsubsection{Routing strategy based on congestion control}

Contra posing unbalanced traffic distribution characteristics, achieving load balancing in the network and optimizing network resource allocation is imperative to prevent or avoid the occurrence of congestion. Multipath routing can allocate traffic to the multiple paths, use more nodes to fulfil the traffic transmission task, achieve load balancing, and make full use of network resources. In addition, multipath routing should consider the current traffic amount demand, if only a small amount of traffic needs to be transported, using multipath routing may cause waste of network resources; second, QoS routing protocols should be able to provide corresponding routing policies for different types of traffic according to their QoS. Based on the above demands, this paper proposes a multiple path congestion control algorithm for multi-service traffic: each satellite stores 2 the route table: shortest path route table (route table1) and backup route table based on congestion control. Routing strategy is shown in Figure \ref{fig:2}, it includes:

\begin{figure}[b]
\centering
\includegraphics[width=0.7\linewidth]{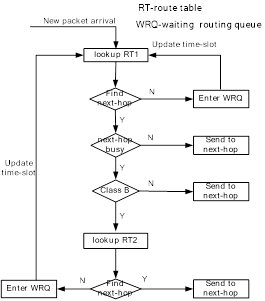}
\caption{On-board routing strategy }
\label{fig:2}

\end{figure}

\begin{itemize}
\item The shortest path route table is established based on virtual topology [6, 7] and calculated offline according to the network topology structure within each time-slot. The connection state and link weights between the network nodes is two main parameters to describe the network topology, for LEO constellation, each satellite can establish connection with 4 neighbor satellites, build two inter-satellite links(ISL) and 2 inter-orbit links(IOL), ISL length can be generally regarded as constant over time, IOL length is changing as the satellite moves. In addition, ISL connection is permanent, while IOL connection is dynamic, IOL connection will turn off when satellites go through the high latitudes, angle of visibility between satellites is too small, and satellites in the cross-seam do not have IOL [2].

\item For satellite traffic arrival rate $\lambda$ set two state thresholds: idle threshold $\alpha$ and busy threshold $\beta$; Satellite nodes keep track of their own traffic arrival rate, when  $\lambda > \beta $  namely determining satellite into busy state, when $\lambda < \alpha $  determining satellite into idle state, defining $\alpha < \lambda < \beta $ as a transition state. Satellite state is monitored, when state change busy/idle occurs, notice the neighbor satellites and routing control center, when neighbor satellite receives "busy" signal, it will reduce traffic to congestion nodes, and the routing control center will remove busy satellite node from the network topology and recalculate the backup route table.

\item When retransmit on-board traffic, first find the next hop according to the shortest path route table, check its status, if the next hop node is idle, send forward the traffic to the next hop satellite; If the next hop node in busy state, then judge the forward direction depending on the traffic class: traffic class A  will be forwarded to the next hop satellites directly ,lookup to backup route table and find a new next-hop for class B, if no next-hop node conform to the conditions, traffic class B will be put into routing waiting queue and wait to be forwarded when route table updated.
\end{itemize}

 \section{The network simulation model}

Setting up satellite communication network simulation model is of great significance for the research of LEO satellite communication system, OPNET is currently the world's most advanced simulation platform for the development and application of network and was voted to be the No. 1 "world-class network simulation software" [8], widely used in network planning, while it was downplayed in the field of national defense [9], it has no standard satellite nodes, routing protocol provided cannot be directly used in satellite networks. This paper set the LEO satellite network routing simulation model based on OPNET simulation modeling mechanism [8, 9], satellite communication network is divided into satellite node domain, user node domain and the routing update process

Figure \ref{fig:3}(a) is the satellite node model, ul\_ant, dl\_ant is uplink and downlink antenna module, u\_link is the uplink wireless receiver module, it receives traffic from users and send it to sat\_mac module, d\_link is the downlink wireless transmitter module for sending the traffic to the user. ISL\_rcv, ISL\_txt are inner-satellite wireless receiver and transmitter, respectively responsible for receiving traffic from other satellites and send forward traffic to other satellites. Sat\_dll module is a dynamic linking layer, which converge the traffic from u\_link or ISL\_rcv together and sent it through the queue module to sat\_router module, or distribute the traffic from sat\_router to ISL\_txt or d\_link. The queue module realizes packet scheduling and traffic rate monitoring function, it sends the traffic to sat\_router module in accordance with PQWRR packet scheduling policy and reports satellite status to sat\_router layer.

Figure  \ref{fig:3}(b) shows the process model of sat\_router module, when traffic from the lower layer arrives, it lookup the route table to find the next hop, if could not find the next-hop, packet will be deposited in waiting routing queue, otherwise check whether the corresponding lower layer transmitter has idle channels, if exist idle channel traffic will be sent to the corresponding transmitter through sat\_dll layer, otherwise deposit packet in waiting sending queue. When module sat\_router monitors the lower layer transmitter has idle channel, it will send packets in sending waiting queue forward to the transmitter; when next time-slot arrives update the route table and look up the route table to find the next hop node for packets in waiting routing queue; if the satellite state (busy/idle) changing is monitored, sat\_router module notifies the neighbor node and the routing control center.

Figure  \ref{fig:3}(c) shows the user node model, it includes modules pk\_generator, pk\_rcv, g\_net, g\_mac, u\_link, d\_link, ul\_ant, dl\_ant, etc. Module pk\_generator is used to simulate the traffic requests, module pk\_rcv is designed for receiving traffic and generate statistical information, module g\_net is responsible for registering the address and location information of the user to node routing control center. Module g\_mac is used to find the current access satellite and monitor wireless transmitter state, when the transmitter has an idle channel, traffic will be connected to satellite via the wireless transmitter. u\_link, d\_link, ul\_ant, dl\_ant is designed to realize the function of wireless transmitter, wireless receiver, uplink antenna and downlink antenna respectively.

\begin{figure}[h]
\centering
\includegraphics[width=0.3\linewidth]{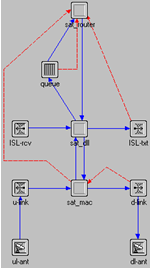}\\
(a) satellite node model\\
\includegraphics[width=0.6\linewidth]{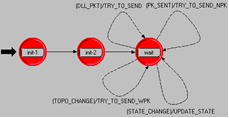}\\
(b) sat\_router module's process model\\
\includegraphics[width=0.6\linewidth]{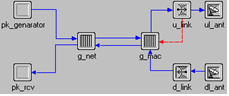}\\
(c) user node model\\
\caption{OPNET modeling node model}
\label{fig:3}
\end{figure}

Packet format is shown in Figure \ref{fig:4}, the field tos identifies type of service, field dst is the mark of destination node address, field next is used to store the next-hop node address, field hop record the hop traffic experienced in the network, field packet represent useful information needs to be sent by the user. 

\begin{figure}[h]
\centering
\includegraphics[width=0.5\linewidth]{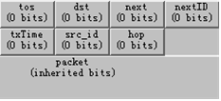}
\caption{Package format}
\label{fig:4}
\end{figure}

Figure \ref{fig:5} shows the routing update process model, and it recalculates the shortest path route table and backup route table when time-slot and satellite node state shift.

\begin{figure}[h]
\centering
\includegraphics[width=0.5\linewidth]{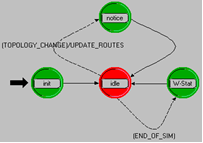}
\caption{The routing update process model}
\label{fig:5}
\end{figure}

\section{The simulation and result analysis}

Simulation scenario is shown in Figure \ref{fig:6}, using iridium constellation model as LEO constellation topology, it has six orbit plane, on each plane distribute 11 satellites, orbit altitude is 780 km, latitude threshold and the minimum elevation are set to be $ 60 \degree $  and  $8.2 \degree$  , orbit file is generated by STK software and then imported to OPNET. Traffic class B is divided into three types- B0, B1, B2, their weights are increasing in turn, proportion of traffic A, B2, B1, B0 is 25\%, packet size is 1000 bits, on-board traffic processing rate is 500 packets/s, and the buffer queue length is 50 packets.

\begin{figure}[t]
\centering
\includegraphics[width=\linewidth]{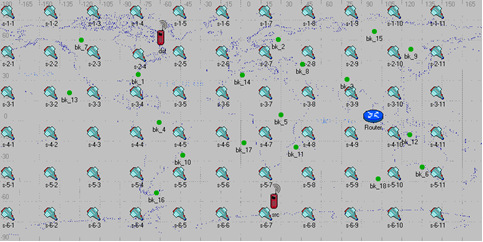}
\caption{The simulation model of the network}
\label{fig:6}
\end{figure}
       
Literature [10] has divided the earth into 12 x 24 cells and predicted the traffic demand in every cell (see figure 7), the literature [11] provides the flow ratio between continents (see table 1). We set traffic background based on these two literatures, using little green dots to simulate the traffic demand of its region and setting the destination address on the basis of traffic ratio between the continents. Source, destination nodes are set respectively in ($56\degree $south latitude, $26\degree $east longitude) and ($65.2\degree $north latitude, $58\degree $ west longitude), total traffic demand between them is 100 packets/s. We set up two simulation scenarios, respectively to validate the performance of PQWRR scheduling algorithm and the composite routing algorithm, each policy is simulated for 30 min.

Figures \ref{fig:8}-\ref{fig:10}  show packet loss status under PQWRR strategy respectively on satellite S-1-1, S-1-4, S-5-4, packet loss on satellite S-1-1 occurs in 11th min and 20th min, on satellite S-1-4, S-5-4 packet loss occurs in 10th min, 4th min respectively, and on-board load is constantly changing as cover domain changing due to satellite movement.

\begin{figure}
\centering
\includegraphics[width=\linewidth]{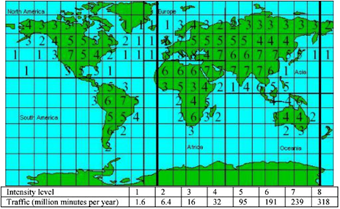}
\caption{Global traffic demand distribution [10]}
\label{fig:7}
\end{figure}

\begin{table}
\centering
\renewcommand{\arraystretch}{1.3}
\begin{tabular}{|c|c|c|c|c|c|c|}
\hline
\multirow{2}{*}{Src}    & \multicolumn{6}{c|}{Dst}                                     \\
\cline{2-7}
                        & 1        & 2        & 3        & 4        & 5      & 6       \\
\hline
1                       & 86.18    & 6.74     & 4.18     & 1.76     & 0.45   & 0.70    \\
\hline
2                       & 25.10    & 55.88    & 13.52    & 1.62     & 2.84   & 1.04    \\
\hline
3                       & 24.04    & 20.89    & 47.74    & 1.15     & 1.75   & 4.43    \\
\hline
4                       & 52.39    & 13.02    & 5.96     & 25.12    & 1.85   & 1.66    \\
\hline
5                       & 25.63    & 43.34    & 17.33    & 3.53     & 7.95   & 2.22    \\
\hline
6                       & 26.48    & 10.58    & 29.22    & 2.11     & 1.49   & 30.12   \\
\hline	
\end{tabular}
\vspace{0.2cm}
\\
1-North America 2-Europe 3-Asia 4-South America 5-Africa 6-Oceania
\caption{Traffic ratio between the continents [11]}
\label{table:1}
\end{table}

At the same time packet loss status on different satellites is significantly different, existing traffic unbalance in time and space, packet loss rate is a function of time and space. Due to PQWRR scheduling strategy, traffic class A has the priority right of preemption for satellite resources, its packet loss rate has remained 0, traffic class B suffers packet loss during the on-board resource shortage, because inner traffic class B using weighted round-robin scheduling, only traffic class B0 with the lowest weight suffers packet loss when congestion is lighter (see Figure \ref{fig:10}), but when congestion get heavier, all the traffic class B suffers packet loss in different degrees. We use the alternative routing path for traffic class B with lower priority to ensure its throughput, route hop under composite strategy is shown in Figure \ref{fig:11}, traffic class A is transmitted on the shortest path always, the hop count of the shortest path changes between 6 and 7, while in the 12th route\_hop\_A jump to 9, this is because time-slot updates during the traffic routing process, route table changes, routing path shifts, the routing hop count increases. When each satellite on the shortest path are relatively free, the two classes of traffic are routed on the shortest path, their hops are same. When congestion occurs, route\_hop\_B increases, in most cases increase of the hop is not much, not greater than 2, but in some moments alternate paths hop is extremely large, the maximum hop of class B comes out to be 14. This result is related to the network state, in one moment a larger number of network nodes are marked as busy, leading to the increase of hop count of alternative paths.

\begin{figure}
\centering
\includegraphics[width=0.5\linewidth]{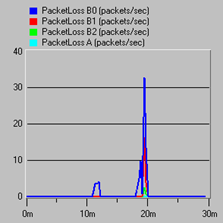}
\caption{Packet loss status under PQWRR strategy on S-1-1}
\label{fig:8}
\end{figure}

\begin{figure}
\centering
\includegraphics[width=0.5\linewidth]{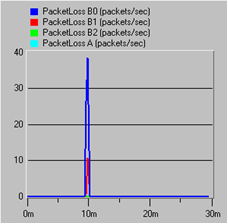}
\caption{Packet loss status under PQWRR strategy on S-1-4}
\label{fig:9}
\end{figure}

\begin{figure}
\centering
\includegraphics[width=0.5\linewidth]{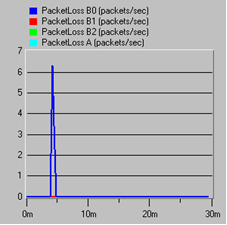}
\caption{Packet loss status under PQWRR strategy on S-5-4}
\label{fig:10}
\end{figure}

\begin{figure}
\centering
\includegraphics[width=0.5\linewidth]{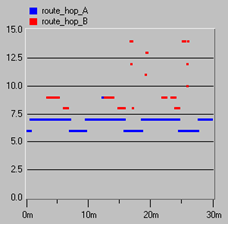}
\caption{Route hop under composite strategy}
\label{fig:11}
\end{figure}

The statistical curve of time delay performance is shown in Figures \ref{fig:12}-\ref{fig:15}, due to the traffic distribution imbalance, at some point vast traffic converge together on the same satellite and lead to network congestion, PQWRR scheduling policy effectively guarantee the ETE delay performance of real-time traffic, it remains around 100ms and delay jitter is lighter. However due to the congestion, the bandwidth of on real-time traffic is preempted by the real-time traffic, QoS performance of class B get worse, its ETE delay and delay jitter get larger (see Figure \ref{fig:12}).

\begin{figure}
\centering
\includegraphics[width=0.5\linewidth]{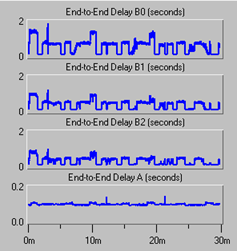}
\caption{ETE delay under PQWRR strategy}
\label{fig:12}
\end{figure}

\begin{figure}
\centering
\includegraphics[width=0.5\linewidth]{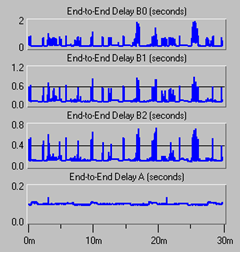}
\caption{ETE delay under composite strategy}
\label{fig:13}
\end{figure}

We can see from Figure \ref{fig:13}, when composite PQWRR scheduling strategy and alternate path, non-real-time traffic choose relatively free satellite as a relay node in the network, the end-to-end delay dramatically reduce , but due to the multipath strategy and the routing hop relatively increase, which cause the severe mutations of ETE delay of class B. While, we found that at some point the time delay of class B has not been improved obviously, this is because although the queuing delay on single satellite reduced, but the routing hop count get larger, times of queued get larger, path length get longer, these lead to the increase of delay.

\begin{figure}
\centering
\includegraphics[width=0.5\linewidth]{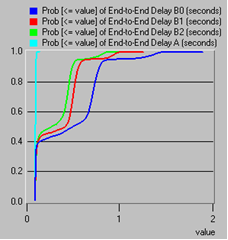}
\caption{CDF of ETE delay under PQWRR strategy}
\label{fig:14}
\end{figure}

\begin{figure}
\centering
\includegraphics[width=0.5\linewidth]{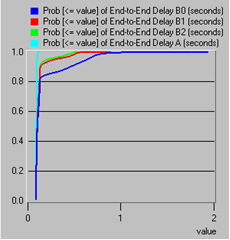}
\caption{CDF of ETE delay under composite strategy}
\label{fig:15}
\end{figure}

The cumulative distribution (CDF) curves of ETE delay are given to describe delay performance of the two routing policies more intuitively, CDF of real-time traffic under two strategies have not much difference. For traffic class B, CDF curve of ETE delay under PQWRR strategy rises slowly, and time delay distribution range is wide, on the contrary, the cumulative distribution curve under composite strategy rises rapidly, delay distribution range is narrow and focused on the low range. Under the PQWRR scheduling strategy, with network congestion traffic with lower priority suffers a long time queuing to get service, a large number of lower priority traffic delay distributes within high delay range, 90\% ETE delay of the traffic class A, B2, B1, B0 respectively less than 102ms, 98ms, 567ms and 790ms. In combination with alternative paths, traffic with low priority get service through idle satellite nodes, which reduces the on-board queuing delay, 90\% ETE delay of traffic class B2, B1, B0 respectively less than 136ms, 145ms and 460ms, ETE delay performance of traffic class B with different weights has been significantly improved, network performance gets optimized obviously.

\begin{figure}
\centering
\includegraphics[width=0.5\linewidth]{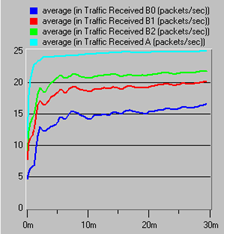}
\caption{Packet throughput under PQWRR strategy}
\label{fig:16}
\end{figure}

\begin{figure}
\centering
\includegraphics[width=0.5\linewidth]{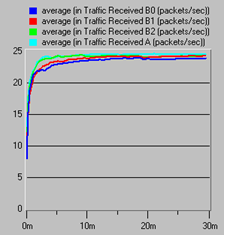}
\caption{Packet throughput under composite strategy}
\label{fig:17}
\end{figure}

Figures \ref{fig:16},\ref{fig:17} describe the throughput characteristics of the two strategies, it can be seen from the figure 16 that when the network congestion occurs, the throughput of the various traffic exists obvious differences in resources shortage. Cause the real-time traffic has preemption priority for bandwidth, its throughput performance obtained powerful guarantee; but non real-time traffic shares the rest of the bandwidth according to their respective weights using the WRR scheduling policy, the throughput performance is not ideal, the lower traffic priority, the worse throughput performance, especially class B0, due to its weak resource competitiveness, its throughput is only about 15 packets/s, throughput rate only reached to 60\%. Compared with pure PQWRR resource allocation strategy, with hybrid routing strategy, due to the effective use of bandwidth resources of the free satellite nodes, non-real-time traffic bandwidth obtains very good compensation, which improves the throughput performance largely, packet throughput has been effectively guaranteed, every type of packet throughput have reached more than 90

\section{Conclusion}
Traffic supported by the future satellite networks must be diversified with high speed, satellite network has to assume a plenty of traffic, but the traffic distribution is extremely unbalanced, satellite constellation topology changes constantly, in some time due to traffic convergence some satellite nodes bear heavy load while the other bear the very small amount of traffic. Contra posing this phenomenon, this paper puts forward a composite routing strategy combining PQWRR scheduling strategy and alternate path routing algorithm together. When traffic compete for network resources, proceeding from the view of differentiated service, first guarantee the time delay, delay jitter, bandwidth, throughput requirements for real-time traffic, then let all levels of non-real-time traffic shares the rest resources according to their own weights. In order to avoid low priority traffic to be "starve" due to its weak competitiveness in the process of resource competition, in the network congestion occurs, adopt the strategy of alternative paths, to shunt low priority traffic to more free satellite nodes.

It alleviates the burden of congestion satellite and compensate at the same time the bandwidth of the low priority traffic. Finally we set up the LEO satellite communication network simulation platform in OPNET, the simulation result shows that the proposed algorithm has the very good performance to ensure the QoS of real-time and non-real-time traffic in aspect of ETE delay, throughput, although under this policy the delay jitter of class B is severe, but as a whole, for traffic class B the ETE delay reduce greatly, throughput improves significantly. But in this paper how to choose state threshold α, β has not been discussed, the selection method will be the focus of our future research.

\section*{Acknowledgment}
We would like to thank the Center for Artificial Intelligence and Robotics (CAIRO), Universiti Teknologi Malaysia (UTM) for providing the research facilities. This research work has been supported.



%

\end{document}